\documentclass[floatfix,twocolumn,bibnotes,prl,amsmath,superscriptaddress,
               preprintnumbers,amssymb,amsfonts,
               letterpaper]{revtex4}

\usepackage{graphicx} 

\begin{document}

\newcommand{\BTO}{\mathrm{BaTiO}_3}
\newcommand{\STO}{\mathrm{SrTiO}_3}
\newcommand{\SRO}{\mathrm{SrRuO}_3}

\title{Large bias-dependent magnetoresistance in all-oxide magnetic tunnel junctions with a ferroelectric barrier}

\author{Nuala M. Caffrey, Thomas Archer, Ivan Rungger and Stefano Sanvito}
\email[corresponding authors: ]{caffreyn@tcd.ie,  sanvitos@tcd.ie}
\affiliation{School of Physics and CRANN, Trinity College, Dublin 2, Ireland}

\date{\today}

\begin{abstract}
All-oxide magnetic tunnel junctions (MTJs) incorporating functional materials as insulating barriers have the potential of becoming the 
founding technology for novel multi-functional devices. We investigate, by first-principles density functional theory, the bias-dependent 
transport properties of an all-oxide SrRuO$_3$/BaTiO$_3$/SrRuO$_3$ MTJ. This incorporates a BaTiO$_3$ barrier which can be found 
either in a non-ferroic or in a ferroelectric state. In such an MTJ not only can the tunneling magnetoresistance reach enormous values, 
but also, for certain voltages, its sign can be changed by altering the barrier electric state. These findings pave the way for a new generation 
of electrically-controlled magnetic sensors.
\end{abstract}

\maketitle


The control of the spin dependent tunneling between two ferromagnetic electrodes separated by an insulating barrier has enabled
enormous advances in many magnetic data storage technologies, in particular since extremely large tunneling magnetoresistance 
(TMR) was measured. The progress in producing magnetic tunnel junctions (MTJs) with large TMR was initially limited by the use 
of amorphous tunnel barriers. The situation however changed after the prediction \cite{PhysRevB.63.054416, PhysRevB.63.220403} 
and subsequently experimental realization \cite{Parkin1,Yuasa} of epitaxial MTJs. Since then, room temperature TMR in excess of 
600\% has been demonstrated in MgO-based devices \cite{ISI:000259011900060}. 


In general, for amorphous barriers the spin polarization of the tunneling current and hence the TMR magnitude, depend solely on the 
electrodes' density of states (DOS) at the Fermi level, $E_\mathrm{F}$ \cite{Julliere}. In contrast, perfectly crystalline tunnel barriers are 
wave-function symmetry selective and make the tunneling process sensitive to their electronic structure. As a result the amplitude and 
even sign of the TMR may depend on the barrier itself. The understanding of such a concept suggests that one can engineer the TMR 
by carefully selecting the insulating barriers to be epitaxially grown on magnetic electrodes. Ferromagnets  \cite{Modera} and ferroelectrics
\cite{Gajek,ISI:000274901100029} are of particular interest as functional barriers.

Ferroelectric materials possess a spontaneous electric polarization whose direction can be switched by an electric field. This makes 
ferroelectric-based MTJs fully multi-functional devices able to respond to both electrical and magnetic stimuli \cite{Gajek,TsymbalNanoLetters}. 
Importantly ferroelectrics can be grown epitaxially on a variety of substrates \cite{EpiFerro} but in particular on other oxides. Since epitaxial 
growth is a pre-requisite for large TMR, the prospect of all-oxide junctions appears particularly attractive. Such a type of MTJ is investigated 
in this Letter. We demonstrate theoretically a huge TMR and more importantly we show that the TMR sign can be reversed with bias, at a critical 
bias which depends on the ferroic state of the barrier. Our results are rationalized in terms the band-structure match between the ferroelectric 
insulator and the ferromagnetic electrodes. 

Density functional theory (DFT) calculations are performed with the local basis set code \textsc{siesta}~\cite{Siesta}. Structural relaxation 
is obtained with the generalized gradient approximation (GGA) of the exchange and correlation functional \cite{PhysRevLett.77.3865}. 
This gives a satisfactory device geometry, but it produces a rather shallow band alignment mainly because of the DFT-GGA gap problem. 
In order to make up for this shortfall the electronic structure used for the transport calculations is that obtained with the atomic self-interaction 
correction (ASIC) scheme \cite{Pemmaraju/Sanvito:2007}, which improves drastically the electronic properties of both bulk BaTiO$_3$~\cite{Tom} 
and SrRuO$_3$~\cite{rondinelli:155107}. Unfortunately the approximate ASIC energy functional is not sufficient to produce good structural 
parameters and in particular the BaTiO$_3$ ferroelectric state cannot be stabilized. This is a current limitation of the method, which otherwise 
has been successful in predicting the electronic properties of oxides \cite{Andrea}. For this reason we perform ASIC transport calculations at 
the GGA relaxed structural parameters. For all the calculations we use a 6$\times$6$\times$1 $k$-point Monkhorst-Pack mesh and a grid 
spacing equivalent to a plane-wave cutoff of 800~eV. 

Electron transport is computed with the \textsc{smeagol} code \cite{Smeagol1,Smeagol2}, which combines the non-equilibrium Green's function 
scheme with DFT. Since \textsc{smeagol} interfaces \textsc{siesta} as the DFT platform, we employ here the same parameters used for the total 
energies calculations. In brief, the total electronic current is given by
\begin{equation}
I^{\sigma}(V) = \frac{e}{h} \int dE\; T^\sigma(E;V)\; [f_L - f_R]\:,
\end{equation}
where $\sigma$ labels the spin ($\uparrow,\downarrow$), $T^\sigma(E;V)$ is the energy dependent transmission coefficient for the bias $V$,
$f_{L/R}$ is the Fermi distribution function evaluated at $E - \mu_{L/R}$ and $\mu_{L/R} = E_{F} \pm \frac{eV}{2}$ is the chemical 
potential of the left/right electrode. If the junction is perfectly translational invariant in the plane orthogonal to the transport direction, 
$T^\sigma$ is obtained by integrating the $\vec{k}$-dependent $T_{\vec{k}}^\sigma$ over the 2D Brillouin zone 
of volume $\Omega_\mathrm{BZ}$,
\begin{equation}
T^\sigma(E, V) = \frac{1}{\Omega_\mathrm{BZ}} \int_\mathrm{BZ} d\vec{k} \; T_{\vec{k}}^\sigma(E;V)\:.
\end{equation}

We initially perform relaxation of bulk $\BTO$ and $\SRO$ under an in plane compressive strain emulating the common epitaxial growth 
on $\STO$. The relaxed cells are then used to construct the transport supercell, which comprises of six $\BTO$ unit cells ($\sim$2.5~nm) 
sandwiched at either side by three $\SRO$ ones. The $\SRO$/$\BTO$ interface is SrO/TiO$_2$, due to the experimentally observed volatility 
of the RuO$_2$ termination \cite{rijnders:505}. We consider two structures. In the first non-ferroic (NFE) structure the atoms are frozen artifically 
in their centro-symmetric positions with the interfacial distance given by an average between the $\BTO$ and $\SRO$ $c$-lattice constant. 
In the second the supercell is further relaxed with respect to the atomic coordinates, resulting in a stable ferroelectric ground state (FE structure).
\begin{figure}[h]
\begin{centering}
\includegraphics*[width=1.0\linewidth]{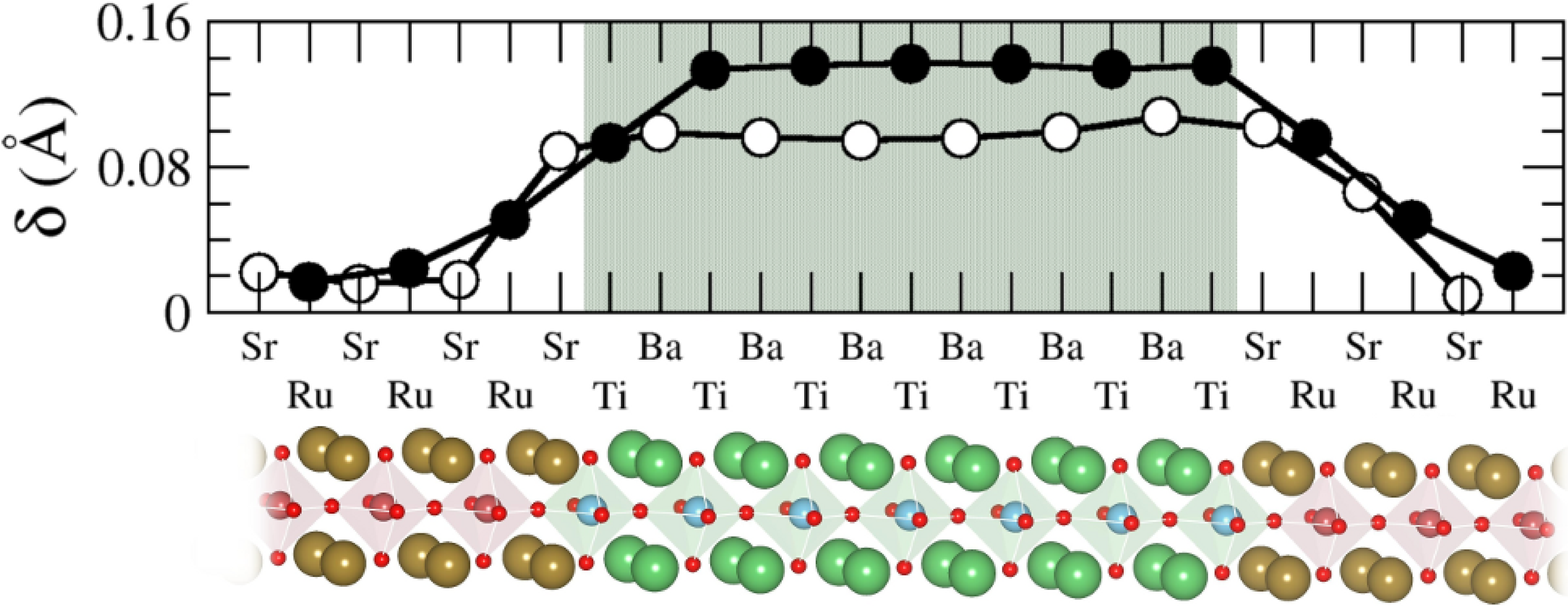}
\caption{\label{fig:displacements}\small{(Color online) Relative atomic displacement for Sr, Ba (open circle) and Ru, Ti (filled circle) 
in the $\SRO/\BTO/\SRO$ MTJ investigated. The displacements are with respect of the O atoms in the same plane, 
$\delta = (z_\mathrm{cation} - z_\mathrm{O})$, with $z$ being stack direction. The device geometry is presented in the 
lower part of the figure.}}
\end{centering}
\end{figure}

The atomic relaxed displacements, $\delta$, with respect to the planar O positions are shown in Fig.~\ref{fig:displacements}. At the center of the 
the $\BTO$ slab Ti displaces by 0.14~\AA, which is significantly smaller than the value of 0.23~\AA\ of bulk BaTiO$_3$ experiencing the same strain. 
Note that GGA overestimates the volume and atomic distortions associated with ferroelectricity in BaTiO$_3$ resulting in a ``super-tetragonal'' 
structure. Such an overestimation, while resulting in a polarization greater than the experimental one, has not a significant qualitative effect on 
our results. The interfacial SrRuO$_3$ layers, as expected, also contribute to the polarization \cite{PhysRevLett.96.107603}.


The symmetry of the electronic bands of both the ferromagnetic electrodes and the insulating spacer dictates the transport properties. 
A wavefunction, whether propagating or evanescent, is described in terms of irreducible representations of the crystal's symmetry group. 
For a cubic space group, the $\Delta_1$ symmetry transforms as a linear combination of $1$, $z$ and $2z^2 - x^2 - y^2$ functions, 
while the $\Delta_5$ as a linear combination of $zx$ and $zy$ (e.g. $p_x$, $p_y$, $d_{xz}$ and $d_{yz}$). Finally the $d_{x^2 - y^2}$ 
and $d_{xy}$ states have $\Delta_2$ and $\Delta_{2^{'}}$ symmetry respectively. Importantly an incident Bloch state in the electrodes can 
couple to a given evanescent state in the insulator, and then sustain a tunneling current, only if the two share the same symmetry. 

\begin{figure}[ht]
\begin{centering}
\includegraphics[width=0.85\linewidth]{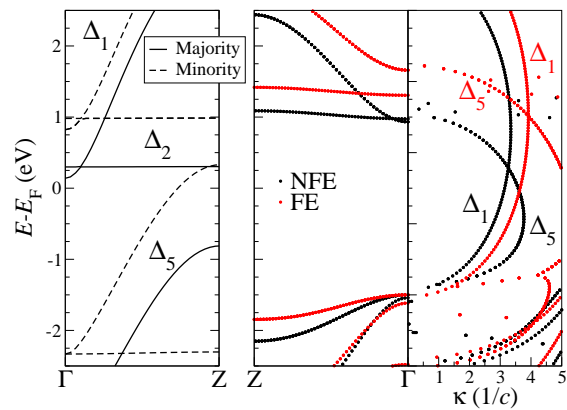}
\caption{\label{fig:sro_bto}\small{(Color online) ASIC-calculated band-structure along the transport direction ($\Gamma\rightarrow Z$) for 
centro-symmetric tetragonal SrRuO$_3$ (left), and BaTiO$_3$ (middle) both in the NFE (green) and FE (red) configuration. The wavefunction 
symmetries of the bands close to $E_\mathrm{F}$ are indicated. The right panel reports the complex band-structure for BaTiO$_3$. The energies 
are aligned with the $E_\mathrm{F}$ of SrRuO$_3$.}}
\end{centering}
\end{figure}

The left panel of Fig.~\ref{fig:sro_bto} shows the SrRuO$_3$ band-structure close to $E_\mathrm{F}$ along the direction of the 
transport.  At $E_\mathrm{F}$ only a doubly-degenerate minority $\Delta_5$ state is available, in contrast to previous DFT calculations, 
where both minority ($\downarrow$) $\Delta_5$ and majority ($\uparrow$) $\Delta_1$ bands were found~\cite{TsymbalNanoLetters}. 
Such a discrepancy is due to the use of the GGA functional in Ref.~\cite{TsymbalNanoLetters}, which underestimates the Ru $d$ manifold
exchange splitting \cite{rondinelli:155107}. Note that a large spin splitting is expected based on point contact Andreev reflection 
experiments \cite{PCAR}.

In the right panel of Fig.~\ref{fig:sro_bto} we plot the $\BTO$ real and complex band-structure. In contrast to MgO, where states 
with $\Delta_5$ symmetry decay significantly faster than those with $\Delta_1$ \cite{PhysRevB.63.054416}, in NFE $\BTO$ the 
$\Delta_1$ and $\Delta_5$ symmetries have comparable decay rates. In particular close to the valence band top the slower 
decay rate is for $\Delta_1$, while the situation is reversed at the conduction band minimum. The enlargement of the bandgap 
associated with the FE order results in an increased decay rate for all the symmetries. The effect is more pronounced for $\Delta_5$ 
close to the top of the valence band where now the $\Delta_1$ symmetry primarily contributes to the tunnel conductance. 

\begin{figure}[ht]
\begin{centering}
\includegraphics[width=0.80\linewidth]{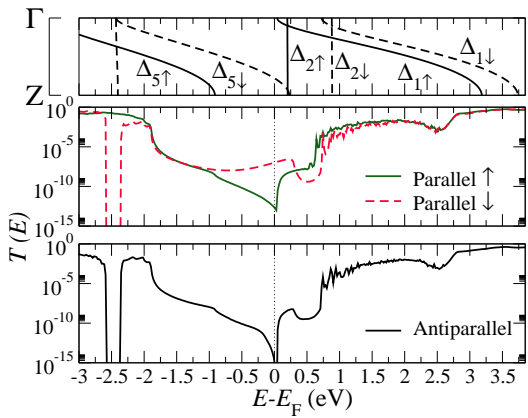}
\caption{\label{fig:para_p_ap}\small{(Color online) Transmission coefficients, $T(E)$, as a function of energy for the NFE structure. 
The middle panel is for the parallel magnetic configuration, while the lower one is for the antiparallel. At the top we report again 
the SrRuO$_3$ band-structure at the $\Gamma$ point of the 2D transverse BZ. The dotted line at 0~eV denotes $E_\mathrm{F}$.}}
\end{centering}
\end{figure}

We begin our analysis of the transport properties from the NFE structure by showing $T(E)$ at zero bias for the parallel (PA) and 
antiparallel (AP) magnetic alignment of the electrodes (Fig.~\ref{fig:para_p_ap}). In the PA configuration $T(E)$ close to $E_\mathrm{F}$ 
is dominated by the minority spin channel. This is expected from the band-structure of $\SRO$, which presents only a 
$\Delta{_5\downarrow}$ band along the transport direction for energies comprised between $-0.8$~eV and +0.1~eV. The minority 
conductance in this energy range is five orders of magnitude larger ($T^\downarrow\sim10^{-7}$) than that for the majority spins. 
For $E>0.1$~eV there is a sharp rise in $T^\uparrow$, due to the $\Delta{_1\uparrow}$ band now contributing to the conductance. 
In the energy window 0.3~eV~$<E<$~0.8~eV there are no minority states available and $T^\downarrow$ drastically drops. A similar 
drop, due to the lack of minority $\SRO$ bands is found at -2.5~eV.  
In the AP configuration the electron transmission occurs between majority (minority) states in the left-hand side electrode and 
minority (majority) in the right-hand side one, so that $T(E)$ for both the spins (identical) is essentially a convolution of those 
for the majority and minority spin channels of the PA state. As a consequence there is a drastic suppression of $T(E)$ in the regions 
-0.8~eV~$<E<0.1$~eV and 0.3~eV~$<E<0.8$~eV, where respectively the $\Delta{_5\downarrow}$ and $\Delta{_1\uparrow}$ 
bands in one electrode are not paired in the other. In particular $T(E_\mathrm{F})$ for the AP configuration is orders of magnitude 
smaller than in the PA one. Note that our discussion is based on the band-structure at the $\Gamma$-point of the 2D 
transverse BZ, for which the decay is the smallest and the transmission the largest. However, also Bloch states with larger 
transverse wave-vector contribute to the transport and produce a residual transmission.

\begin{figure}[ht]
\begin{centering}
\includegraphics[width=0.90\linewidth]{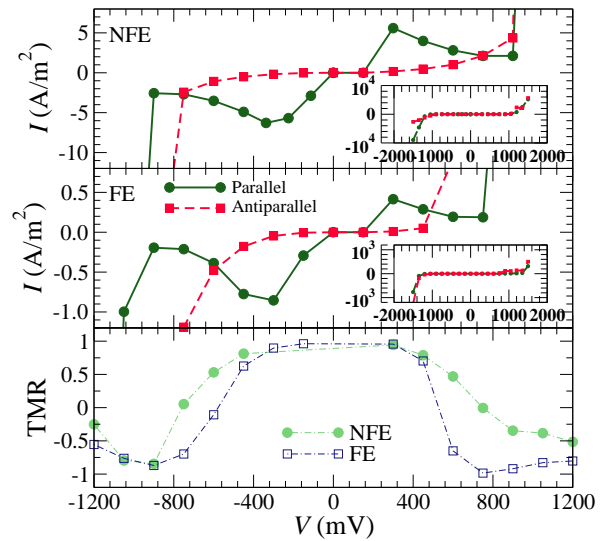}
\caption{\label{fig:current}\small{(Color online) Total current per unit area, $I$, as function of voltage, $V$, for the NFE (top panel) and FE 
(middle panel) structures. In the bottom panel we present the TMR as a function of voltage for both the geometries. In the insets the $I$-$V$
is presented over a larger current range (the units are the same as in the main figure). Note that at the on-set of the $\BTO$ conduction
and valence bands the current increases by three orders of magnitude over its low bias value.}}
\end{centering}
\end{figure}

The spin-polarized current for both the PA and AP configurations and for both the NFE (top panel) and FE (middle panel) structures 
are shown in Fig.~\ref{fig:current}, where we focus on the low voltage region in which the current is due entirely to tunneling (the broader 
$I$-$V$ are displayed in the insets). 
The most distinctive feature emerging from the $I$-$V$ curves is the presence of negative differential resistances (NDR) for the PA 
alignment, originating from the movement of the $\Delta{_1\uparrow}$ band-edge with $V$. Because of the NDR the relative magnitude 
of the current for the parallel ($I^\mathrm{PA}$) and antiparallel alignment ($I^\mathrm{AP}$) can be reversed, i.e. the TMR changes 
sign with $V$. This is demonstrated in the lower panel of Fig.~\ref{fig:current}, where we present the ``pessimistic'' TMR ratio, 
TMR=$(I^\mathrm{PA}-I^\mathrm{AP})/(I^\mathrm{PA}+I^\mathrm{AP})$, as a function of bias. Clearly TMR sign inversion is observed for 
both the NFE and the FE junctions for voltages in the range 0.7-0.9~V. Furthermore for $V\sim0.7$~V the TMR for the NFE junction 
is positive, while that of the FE one is negative, meaning that subtle changes in the barrier electronic structure, such as those induced by 
ferroelectricity, are sufficient to change the sign of the TMR. Note also that the TMR values reported here are actually extremely large. 
For instance for both the NFE and FE junctions and voltages $|V|<0.4$~V the optimistic TMR [$(I^\mathrm{PA}-I^\mathrm{AP})/I^\mathrm{AP}$] 
is around 5,000\%. 

\begin{figure}[ht]
\begin{centering}
\includegraphics[width=1.0\linewidth]{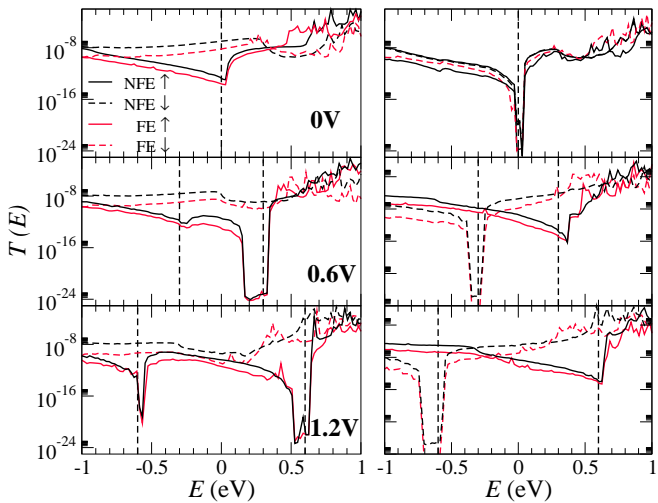}
\caption{\label{fig:fe_transmission_bias}\small{(Color online) Transmission coefficient, $T(E;V)$, as a function of energy and for different
bias voltages. The left panels are for the parallel configuration and the right ones for the antiparallel. In the same figure we report data for both
the NFE structure ($\uparrow$ black solid line, $\downarrow$ black dashed line) and for the PE one ($\uparrow$ red solid line, 
$\downarrow$ red dashed line). The vertical dashed lines mark the borders of the bias window.}}
\end{centering}
\end{figure}

The $I$-$V$ curve can be rationalized by looking at the dependence of $T(E)$ on the bias \cite{IvanTMR}, which is presented in 
Fig.~\ref{fig:fe_transmission_bias}. This is mainly determined by the shift of the electrodes' $\Delta_1$ and $\Delta_5$ band-edges 
with $V$: for positive voltage the band-structure of the left electrode is shifted by $+eV/2$ ($e$ is the electron charge) and that of the right one 
by $-eV/2$. At a given energy a large $T$ is found only if a band of the same symmetry and spin is found in both electrodes at that energy. 
For PA alignment at $V=0$, the minority spins dominate the transmission up to $0.3$~eV, after which one encounters the 
$\Delta_5^\downarrow$ upper band-edge and $T^\downarrow$ is drastically reduced. As $V$ is applied, the $\Delta_5^\downarrow$ band edge 
is shifted to lower energies in the right electrode (for $V>0$), so that for $V=0.6$~V the high transmission region extends only up to 
$E_\mathrm{F}$, and for $V=1.2$~V it extends only up to $E_\mathrm{F}-0.3$~eV. This is the origin of the NDR found for the PA alignment. In 
contrast for the AP configuration $T$ is small for energies below 0.1~eV after which it drastically increases because of the $\Delta_1$ conduction 
bands (see Fig. \ref{fig:para_p_ap}). With increasing $V$ the $\Delta_1$ band in the right electrode is shifted to lower energies, so that 
there is a rather large transmission inside the bias window, and eventually the AP current therefore becomes larger than the PA one. This 
results in the TMR sign change at about 0.7-0.9~V.  

The main effect of the ferroelectric order on the transport is an increase of the $\BTO$ band-gap, i.e. an increase of the $\Delta_1$ and 
$\Delta_5$ decay coefficients (see Fig.~\ref{fig:sro_bto}). In particular, states with $\Delta_5$ symmetry decay significantly faster in the 
FE MTJ with respect to the NFE one. This results in a global reduction of the transmission although other general features remain rather 
similar in the two cases. The comparison between $T(E;V)$ for the FE and NFE junctions is also presented 
in Fig.~\ref{fig:fe_transmission_bias}. Below $E_\mathrm{F}$ one may note a substantial reduction of the transmission when going from 
NFE to FE for both PA and AP alignment as a consequence of the increased $\Delta_5$ decay rate. 

In conclusion, we have demonstrated huge TMR in an all-oxide ferroelectric MTJ, the sign of which can be inverted as the 
applied bias increases. Furthermore the sign inversion occurs at different voltages for different ferroic states of the barrier. Our finite-bias 
results are explained in term of the electrodes and the barrier band-structures. The possibility to control the TMR by 
manipulating the ferroic state of the barrier in an MTJ opens a potential avenue for the electrical control of magnetic devices. 

This work is supported by Science Foundation of Ireland (Grants No. 07/IN.1/I945), by CRANN and by the EU FP7 ATHENA. Computational resources 
have been provided by the HEA IITAC project managed by TCHPC and by ICHEC.

\bibliographystyle{apsrev}


\end{document}